\documentclass{revtex4}

\usepackage{epsfig}
\usepackage{float}
\usepackage{graphicx}    
\usepackage{dcolumn}     
\usepackage{bm}          
\usepackage{subfigure}   
\usepackage{citesort} 

\begin{document}

\title{Inverse Statistics for Stocks and Markets}
\author{A.\ Johansen}
\email{anders-johansen@get2net.dk}
\affiliation{Teglg\aa rdsvej 119, DK-3050 Humleb\ae k, Denmark}
\author{I.\ Simonsen}
\email{Ingve.Simonsen@phys.ntnu.no}
\affiliation{Department of Physics, NTNU, NO-7491 Trondheim,
  Norway} 
\author{M.H.\ Jensen} 
\email{mhjensen@nbi.dk}
\affiliation{Niels Bohr Institute, Blegdamsvej 17,
  DK-2100 Copenhagen {\O}, Denmark}

\date{\today}

 \begin{abstract}
   In recent publications, the authors have considered inverse
   statistics of the Dow Jones Industrial Averaged
   (DJIA)~\protect\cite{optihori,gainloss,gainloss2}.  Specifically,
   we argued that the natural candidate for such statistics is the
   investment horizons distribution. This is the distribution of
   waiting times needed to achieve a predefined level of return
   obtained from detrended historic asset prices. Such a distribution
   typically goes through a maximum at a time coined the {\em optimal
     investment horizon}, $\tau^*_\rho$, which defines the most likely
   waiting time for obtaining a given return $\rho$. By considering
   equal positive and negative levels of return, we reported
   in~\cite{gainloss,gainloss2} on a quantitative gain/loss asymmetry
   most pronounced for short horizons. In the present paper, this
   gain/loss asymmetry is re-visited for 2/3 of the individual stocks
   presently in the DJIA. We show that this gain/loss asymmetry
   established for the DJIA surprisingly is {\em not} present in the
   time series of the individual stocks. The most reasonable
   explanation for this fact is that the gain/loss asymmetry observed
   in the DJIA as well as in the SP500 and Nasdaq are due to movements
   in the market as a whole, {\it i.e.}, cooperative cascade processes
   (or ``synchronization'') which disappear in the inverse statistics of
   the individual stocks.
\end{abstract}



 \maketitle


\section{Introduction}

What drives prices? This question has been studied for centuries with
quantitative theories dating back at least to
Bachelier~\cite{Bachelier}, who proposed the random walk hypothesis
for price trajectories. As prices in general do not become negative,
economist later realized that a more realistic framework was obtained
by assuming the random walk hypothesis for the logarithm of the
price~\cite{Samuelson}. This has made relative returns the prime focus
of financial investigation with only a few exceptions, such as hedge
funds focusing on absolute returns, and benchmarking are almost
exclusively done by the financial community by comparing relative
returns with respect to a fixed time interval.

Within the current economic paradigm --- the Efficient Market
Hypothesis (EMH)~\cite{Fama} --- the idea of randomly fluctuating
prices in the absence of new hard information has been re-formulated
in the framework of hard-working rational traders with complete
knowledge of all available information whose continuing effort more or
less instantaneously removes any imbalance in prices due to past
differences in the expectations of traders. In short, the EMH states
that current prices reflect all available information about the priced
commodity, {\it i.e.}, all available information is at any given
instant already priced in by the market and any change in prices can
only be due to the revelation of new information. In other words,
there is no free lunch.

%
Hence, the EMH claims that information drives prices.  Unfortunately,
this just leave us with another question, namely how to price in the
available information\footnote{As the efficient market hypothesis only
  speaks about the relation between prices and available information
  it as a consequence can only be tested jointly with some
  asset-pricing model. This introduces a ambiguity in how to interpret
  anomalous behaviour of returns: Is the market inefficient to some
  extent or do we have a bad pricing model? This is in fact a very
  serious problem, since it makes it very difficult or impossible to
  ``prove'' that the market is efficient or inefficient and to some
  extent turns the entire question into a matter of
  belief~\protect\cite{thesis}.}? In the case of a stock, this must be
done by considering how the available information affects {\it future
  earnings of the company}. This obviously (again) introduces some
ambiguity as not only do peoples expectations to a largely
unpredictable future differ, but so do their strategies with respect to
for example investment horizons, {\it i.e.} how long they intend to
hold their investment before taking any profit, and how large a risk
they are willing to take over that time period.

In order to qualify different pricing models {\it etc.}, the financial industry
has performed many statistical studies establishing a number of so-called 
stylized facts~\cite{Book:Bouchaud-2000,Book:Mantegna-2000,Book:Hull-2000} 
as well as benchmarking for the performance of various financial
instruments with respect to investment returns and in its complement, risk 
taking. Due to this focus on returns and risk, most financial studies 
essentially amount to measuring two-point correlations in one way or another, 
most commonly done by studying the distribution of returns calculated 
over some pre-defined fixed time period 
$\Delta t$~\cite{Book:Bouchaud-2000,Book:Mantegna-2000,Book:Hull-2000}.

Empirically it has been established, that for not too long time
intervals $\Delta t$'s, say less than a week, the return distributions
are characterized by ``fat
tails''~\cite{Book:Bouchaud-2000,Book:Mantegna-2000,Book:Hull-2000}.
Fat tails of a distribution refers to a much larger probability for
large price changes than what is to be expected from the random walk
or Gaussian hypothesis. As $\Delta t$ reaches longer time scales, the
distribution of returns gradually converges to a Gaussian
distribution.  However, no real consensus regarding the exact
quantitative nature of this distribution exist. 

>From the point of view of the present authors, a more reasonable
answer to the question of what drives prices, besides the influence of
interest rates and other macroscopic economic factors as well as
quarterly earnings reports {\it etc.} of the company in question, is
{\it human psychology} with all its facets of bounded rationality,
limited information, personal beliefs {\it etc.}  In accordance with
this view, the economics and financial communities have in the past
5--10 years witnessed an increased interest in what has been coined
``Behaviourial Economics''. A prime example of such a study is the
book by R.J.\ Schiller entitled ``Irrational Exuberance''\footnote{The
  title is coined from the label that A.  Greenspan put on the
  development of the stock markets in a speech on December 5, 1996.},
published in 2000, but written {\it before} the crash of April of this
year~\cite{Schiller}. This book gives numerous examples of both
present (at that time) and past events where the price of stocks, and
more relevant, the P/E (price-earning-ratio) has more or less exploded
based on little more than vague notions of a ``new economic
era''\footnote{Even the terminology characterizing the economy prior
  to for example the bubbles preceding the crashes of 1929 and 2000
  was more or less the same~\protect\cite{nascrash}.} and the only
real difference comes with the specific sector driving the bubble
prior to the crashes seen in the past 150 years or so.

If we focus on the US stock market, then in the 1860s and 1870s it was
rail road era, in the 1920s it was utilities, semi-conductors
(``Tektronix'') influenced the development in 1950s and 1960s, in the
1980s it was a general public investment boom driven by liberalization
of the financial markets and, most recently, in second half of the
1990s it was first an emerging market boom and then the notorious
information technology bubble. Most notable are the ``explosions'' in
the P/E in the 1920s and in the period from early 1980s to 2000.
During these periods, the P/E of the SP500 went from about $5$ to $32$
and from about $7$ to $45$, respectively\footnote{The average P/E is
  today around $16$ for US stocks, which is still the highest in
  western markets.}.  This corresponds to a more than six-fold
increase in little less that one and two decades for the two
respective periods.

After the October 1987 crash, R.J.\ Schiller asked a number of
institutional and individual investors the following question: ``Which
of the following better describes your theory of the decline(s): {\it
  (i)} a theory about investor psychology, or {\it (ii)} a theory
about fundamentals such as profits and interest rates?'' The answers
revealed that about 68\% of the institutional investors and 64\% of
the individual investors picked the former alternative. From this
R.J.\ Schiller concluded that: ``It appears that the stock market
crash had substantially to do with a {\it psychological feedback loop}
among the general investing public from price declines to selling and
thus further selling, and so forth''.  This conclusion was in
accordance with his investigation of the news stories published
immediately prior to the crash of October 1929 where also no specific
event or events could be singled out as responsible for the crash.

>From the point of view of human psychology, the financial communities
focus on returns does not make as much sense as one would like. The
obvious answer to the question: ``What return would you like to see a
year from now?'', is of course --- ``As large as possible!'' --- or,
if the question concerned risk --- ``As small as possible!''. A more
natural investment strategy is thus to settle for some return level
and/or maximum risk and then ask how long one must wait in order to
achieve this. But how is one to determine {\it a priori} this
investment horizon in an utterly unpredictable world with respect to
the future some months away?  Even if one solely focus on macroscopic
fundamentals as well as those of the company in question, the
predictability of future earnings is a very difficult question. As
J.P. Bouchaud said in a lecture in Granada of February 2005: ``I do not
know how to price even my own company within a factor of two or
perhaps even three though I know everything to be known about it!''

Another, more philosophical problem, with standard models of financial
markets is the following: If one, as these models do, only assumes
that the market participants are purely selfish individuals, who
optimize their own utility function through fixed time contracts with
other nominally identical individuals, then, despite the achievement
of mutual benefits, the term ``contract'' would not be defined in a
general context. Why is this so?

Because, a general and lasting definition of the term ``contract''
requires long-term (meaning much longer than an ordinary human time
span) institutions, which can only be upheld through non-selfish
behavior. Legal institutions, for example, have a life time of
centuries and hence cannot be put to work within a framework of
selfish individuals who's time horizon can't possibly be longer that
there own life span. Hence, most models of the financial markets are
lacking a very essential ingredient, which one usually denote by the
general word ``culture''. In conclusion, in our opinion most models of
the financial markets lack ``psychology and culture''.

\begin{figure}[t]
  \centering
  \includegraphics*[height=9.5cm,width=13.5cm]{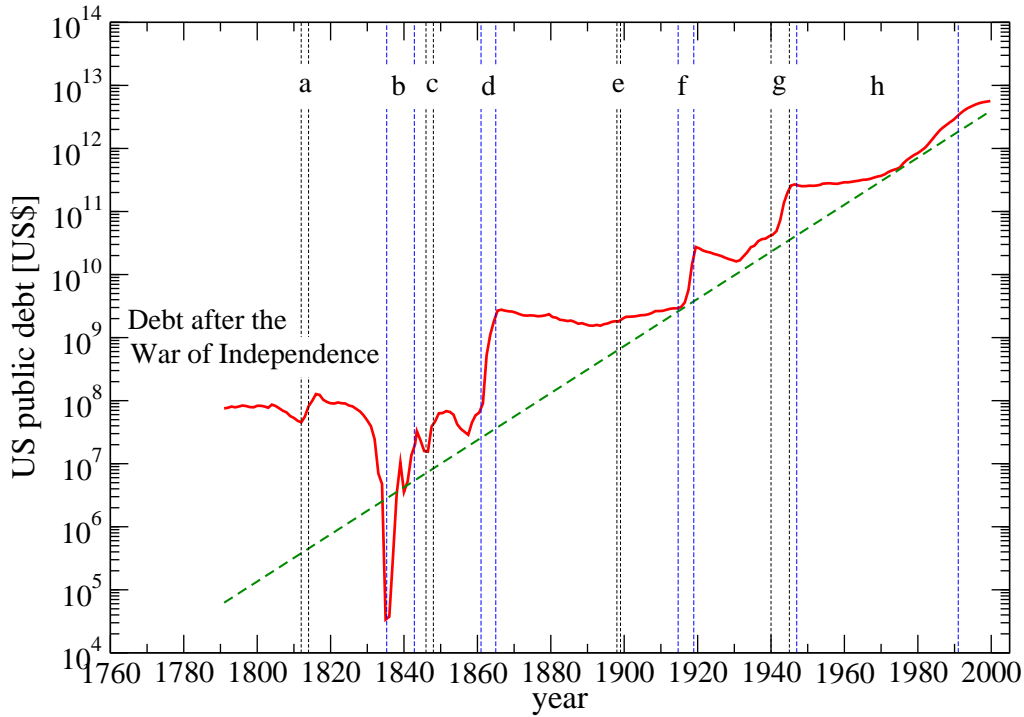}
  \caption{Graph of the historical US public debt from 1791 till 2000.
    The dashed diagonal line represents an exponential function
    corresponding to an average growth rate of about $8.6\%$. Some
    historic events are marked by dashed vertical lines in the figure.
    They are: (a) the 1812 war~(1812--1814); (b) the second war with
    the Seminole Indians~(1835--1842); (c) The Mexican-American War
    (1846--1848); (d) The Civil War~(1861--1865); (e) The Spanish
    American War~(1898); (f) The First World War~(1914--1918); (g) The
    Second World War (1940--1945) (h) The Cold War~(1947--1991).}
  \label{publicdebt} 
\end{figure}

This does not mean, however, that one cannot ask what drives the
economy and the stock market on longer time scales, {\it i.e.} over
decades, and what behaviour the economy exhibits over these time
scales! An example of such a study is presented on
Fig.~\ref{publicdebt}.  Here, the historical public debt of the USA is
shown together with a labeling of the most significant historic
events, mostly military conflicts, involving the US during the same
period.  Notice that the Korean War and it's formal UN-forces did not
represent any significant increase in US public debt. However, the
so-called Cold War between the USA and the USSR most certainly did.
It is clear that the over-all rise in the public debt is exponential
driven by large ``bumps'' signifying rapid increases in public debt.

It is striking, on a qualitative level, that the origin of these large
increases in US public debt is simply related to war with two major
exceptions; the purchase of the Louisiana Territory from Napoleon in
1803 on the unauthorized initiative of the US-ambassador in Paris, and
the Keynesian (in the meaning massive public investments) ``New Deal''
of Roosevelt in the 1930s. Note, that the logarithmic scale is the
reason why the ``bumps'' belonging to the purchase of the Louisiana
Territory and the Spanish War of 1898 do not ``stand out'' in
Fig.~\ref{publicdebt}.

\begin{figure}[t]
  \centering
  \subfigure[DJIA 1810--1902]{
    \includegraphics*[width=0.45\textwidth,height=0.45\textwidth]{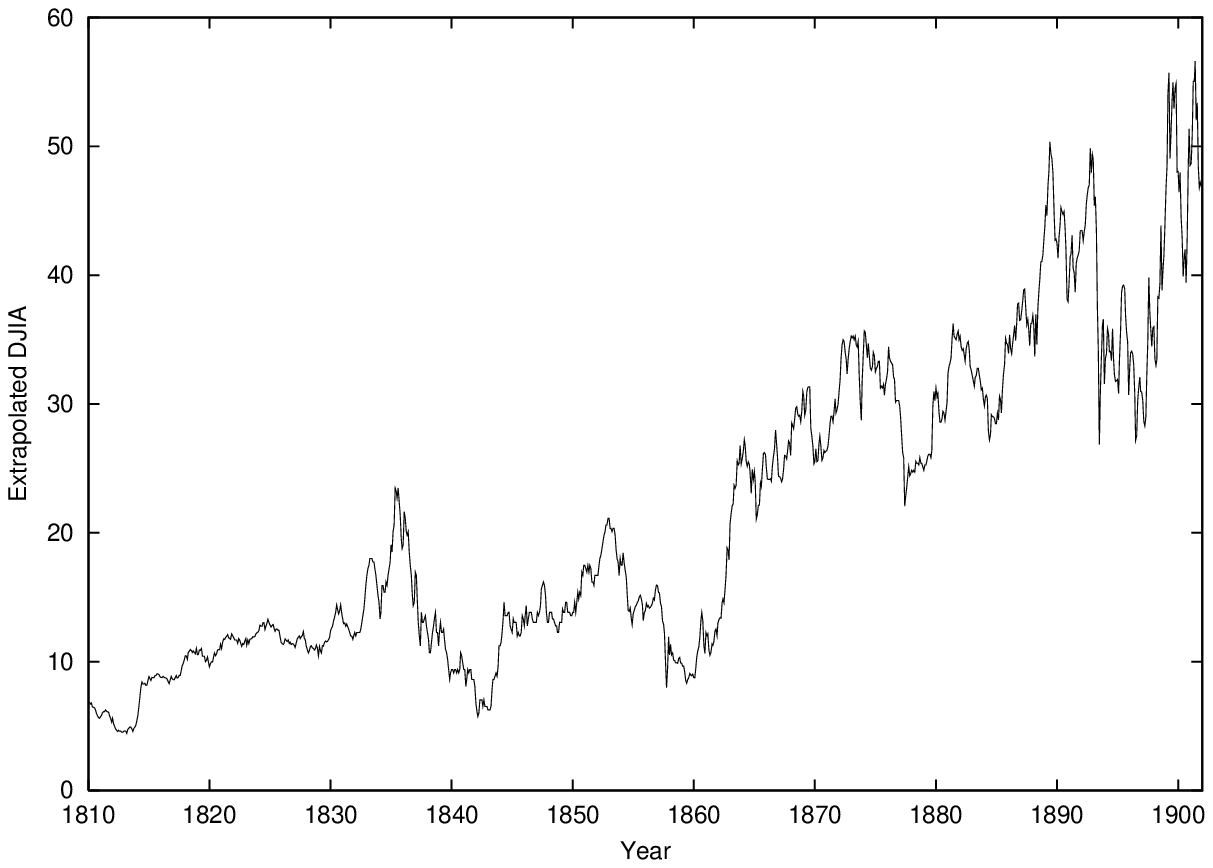} }\quad
  \subfigure[DJIA 1900--1950]{
    \includegraphics*[width=0.45\textwidth,height=0.45\textwidth]{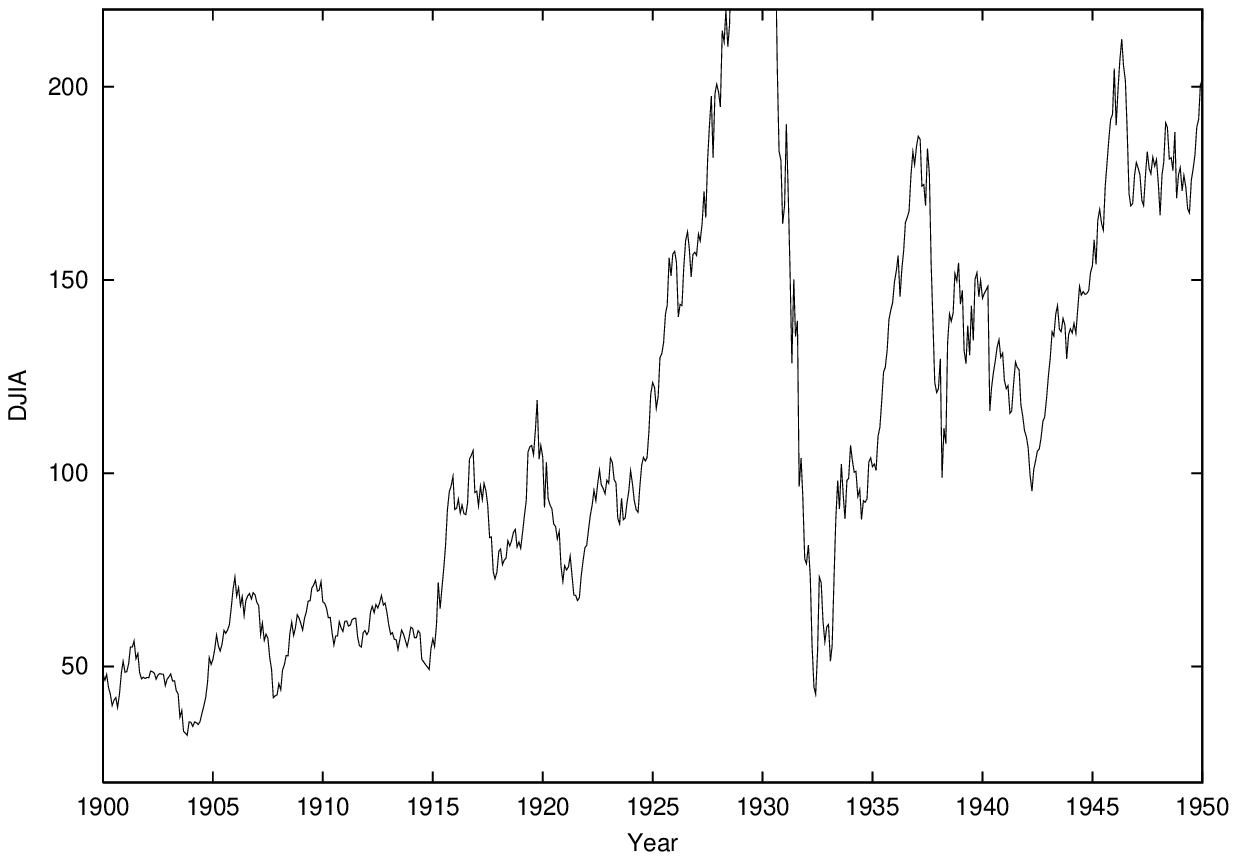} } \\*[0.5cm] 
  \subfigure[DJIA 1896--2001 (incl. detrended data)]{
    \includegraphics*[width=0.95\textwidth,height=0.5\textwidth]{{djia}} } 

  \caption{Historic DJIA data: (a) 1810--1902; (b) 1900--1950; (c)
    1896--2001 including the (wavelet) detrended data series,
    $\tilde{s}(t)$, analyzed below (cf.\ Ref.~\protect\cite{optihori}
    for further details).  }
  \label{djiahistory} 
\end{figure}

If one compares this figure with one of the US stock market, say, the
DJIA (Figs.~\ref{djiahistory}) one clearly sees that on a qualitative
level the rises in the public debt, due to wars and New Deal, are
followed by steep rises in the stock market with one big exception,
namely the bubble of the 1920s. In hind-sight, this may not be so
surprising, since increases in public debt normally means large public
spending which funnels large amounts into the private sector. However,
one should note that the average growth rate of the US public debt is
about $8.6\%$, see Fig.~\ref{publicdebt}. This should be compared with
that of the DJIA until the 1950s, which is about $2.5\%$.  What {\it
  is} surprising is that most time periods with peace exhibit a
significant decline in the public debt, most notably the period after
the war of 1812 until the second war with the Seminole Indians
(1835--1842), as well as modest growth in the DJIA. In the first half
of the 19th century, the US public debt dropped to a meager
US\$~33.733 in 1835 from US\$~75.463.477 in 1791\footnote{In 1790, the
  Federal Government declared that it was redeeming the Scrip Money
  that was issued during the Revolutionary War in the amount of
  US\$~80.000.000.}.
In conclusion with respect to long-term growth in the stock market,
public spending, especially in the case of war, has played a {\em
  very} significant role in the long-term growth of the US economy and
hence of the DJIA.

In the next sections, we will turn to the subject of optimal
investments horizon distributions for both individual stocks as well
as markets. We will do this in order to qualify an answer to the
previous questions re-formulated as ``what a return is reasonable to
expect in $x$ days from now'' for both individual stocks as well as
markets by considering the gain distribution for pre-defined return
levels. In order to quantify the risk from historical data, we also
consider the corresponding loss distribution\footnote{This can in fact
  also be turned into a gain distribution by ``shorting''.}.

The remaining part of this paper is organized as follows. In the next
section, we turn to the short-term behaviour of three US stock
markets, namely the DJIA, the SP500 and the Nasdaq. Here, we will
re-introduce a conceptually new framework for the analysis of
short-term (in the sense of days and weeks) price fluctuations. We
call this tool of analysis for {\it inverse statistics} as we fix the
return level and let time float, and not {\it vice versa}, as in the
usual return statistics. (It is worth noting that an analysis, where
{\it both} return level and time floats conditioned on a stable trend
in either direction has been published by the first
author~\cite{outl1,outl2}). First, we re-visit previous results for
the DJIA as well as presenting new results for the SP500 and the
Nasdaq index, showing that the distribution of time horizons needed to
obtain a specified return can be excellently parameterized by a
generalized gamma-distribution. Such distributions are well-known from
various first-passage problem~\cite{Redner}, but the quality of this
parametrization is nevertheless surprising considering the nature of
the data. We then turn to the previously found {\it gain/loss
  asymmetry} for the DJIA~\cite{gainloss} and show that a similar
asymmetry is present in both the SP500 and Nasdaq.  In the third
section, we turn to the use of the inverse statistics on the single
stocks that are included in the DJIA in order to further investigate
the origin of this gain/loss asymmetry. Surprisingly, we find that the
{\it gain/loss asymmetry} obtained for the index vanishes for the
individual stocks.  The last section concludes arguing that the
gain/loss asymmetry found in the index comes from a cooperative
cascade through the various sectors of the economy represented in the
DJIA.

\section{Previous Work}

In resent publications~\cite{optihori,gainloss,gainloss2,invfx}, the
present authors have proposed to invert the standard
return-distribution problem and instead study the probability
distribution of waiting times needed to reach a {\em fixed level} of
return $\rho$ for the first time (see also Ref.~\cite{mogens}).  As
mentioned previously, this is in the literature known as the ``first
passage time''-problem~\cite{Redner} and the solution is known
analytically for a Brownian motion as the Gamma-distribution
\begin{eqnarray}
p(t) = \left|a\right|\frac{\exp(-a^2/t)}{\sqrt{\pi}t^{3/2}},
\end{eqnarray}
(with $a\propto \rho$), 
 where one for large (waiting) times recovers the 
well-known first return probability for a random walk $p(t) \sim t^{-3/2}$.

Historical financial time series such as the DJIA, SP500 and Nasdaq
possesses an (often close to exponential, see however~\cite{growth})
positive drift over long time scales due to the overall growth of the
economy modulated with times of recession, wars {\it etc.}  If such a
drift is present in the analyzed time series, one can obviously not
compare directly the empirical probability distribution for positive
and negative levels of return. As the focus of the present paper will
be on such a comparison, we must eliminate or at least significantly
reduce the effect of this drift. One possibility for detrending the
data is to use so-called deflated asset prices, but such prices are
calculated using certain economic assumptions, which we as physicists
are naturally suspicious of.

In the present study as well as in those of~\cite{optihori,gainloss},
we have instead chosen to remove the drift based on the use of
wavelets~\cite{Book:NR-1992,AWC}, which has the advantages of being
non-parametric. This technique has been described in detail
elsewhere~\cite{optihori} and for the present purpose, it suffices to
say that this wavelet technique enables a separation of the original
time series $s(t)$ into a short scale (detrended) time series
$\tilde{s}(t)$ and a (long time-scale) drift term $d(t)$ so that
$s(t)=\tilde{s}(t)+d(t)$ (cf. Fig.~\ref{djiahistory}(c)).

Based on $\tilde{s}(t)$ for some historical time period of the DJIA,
the empirical investment horizon distributions, $p(\tau_\rho)$, needed
to obtain a pre-defined return level $\rho$ {\it for the first time}
can easily be calculated for different $\rho$'s. As $\tilde{s}(t)$ is
stationary over time scales beyond that of the applied wavelet (for a
time larger than say 1000 days) it is straightforward to compare
positive and negative levels of return.

As the empirical logarithmic stock price process is known not to be Brownian, 
we have suggested to use a generalized (shifted) Gamma distribution
\begin{eqnarray}
    \label{fit-func}
    p(t) &=&
    \frac{\nu}{\Gamma\left(\frac{\alpha}{\nu}\right)}\,
    \frac{\left|\beta\right|^{2\alpha}}{(t+t_0)^{\alpha+1} }
    \exp\left\{
          -\left(\frac{\beta^2}{t+t_0}\right)^{\nu} 
        \,\right\},
\end{eqnarray} 
where the reasons behind $t_0$ are purely technical\footnote{Its actual value 
depends on possible short-scale drift which may in part be due to the fact that
we are using the daily close. Ideally, one should use some measure of {\it all}
prices during the day, but it's not obvious how to define such a representative
price for the entire trading day. We are aware that the mid price often is used 
as proxy but it is not obvious that this measure is to be preferred as 
relatively large price changes tend to occur in the beginning and end of the 
trading day.}.

\begin{figure}[t]
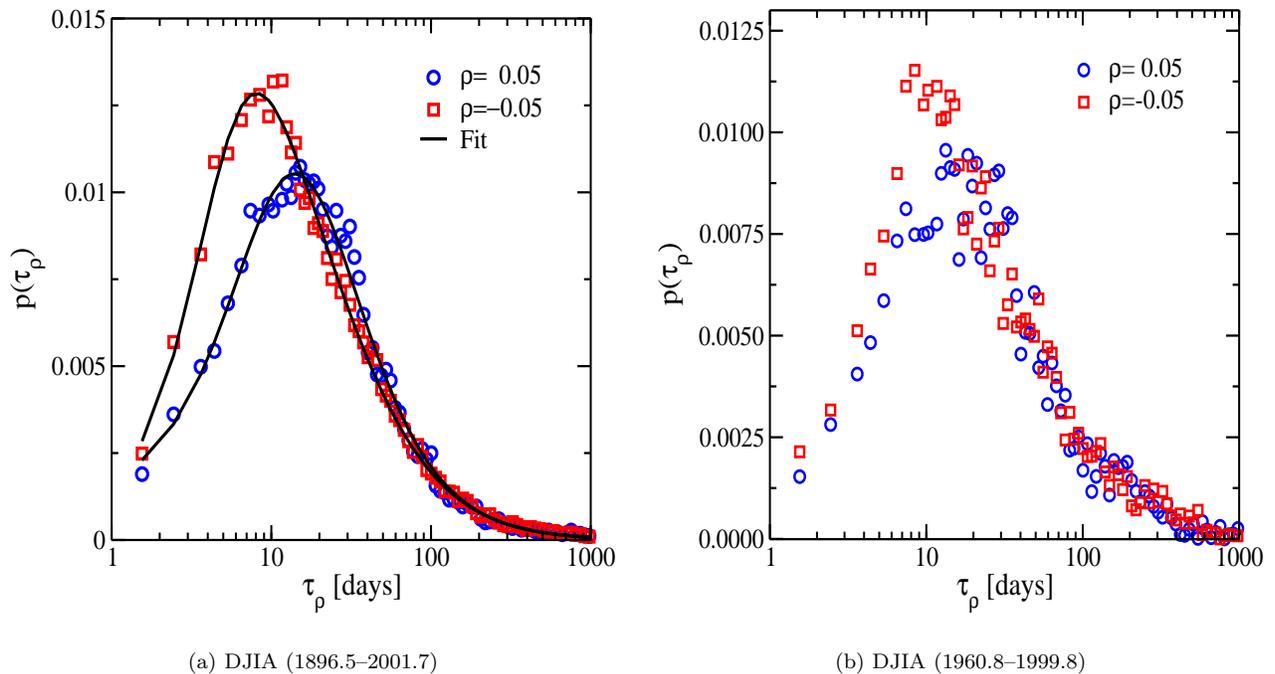

  \centering
  \subfigure[DJIA (1896.5--2001.7)]{
    \includegraphics*[width=0.45\textwidth,height=0.45\textwidth]{djia_r5-pdf} }\quad
  \subfigure[DJIA (1960.8--1999.8)]{
    \includegraphics*[width=0.45\textwidth,height=0.45\textwidth]{djia_short} } 

  \caption{Inverse statistics for detrended closing prices (open
    symbols) of the DJIA for the time periods indicated. For all cases
    the return levels used were $\left|\rho\right|=0.05$. The solid
    lines represent the best fits of Eq.~((\protect\ref{fit-func}) to
    the empirical data with the parameters indicated below; (a) DJIA
    (1896.5--2001.7): $\alpha \approx 0.51$, $\beta \approx 5.23$,
    $\nu \approx 0.68$ and $t_0 \approx 0.42$ (loss distribution);
    $\alpha \approx 0.51$, $\beta \approx 4.53$, $\nu \approx 2.13$
    and $t_0 \approx 10.1$ (gain distribution); (b) Same as
    Fig.~\protect\ref{InvDist_DJIA}(a), but for a shorter time period
    (1960.8--1999.8).  Note that the tail exponents $\alpha+1$ are
    very close to the ``random walk value'' of $3/2$ for all
    distributions.  }
   \label{InvDist_DJIA} 
\end{figure}

\begin{figure}[t]
  \centering
  \subfigure[SP500 (1940.9--2000.3)]{
    \includegraphics*[width=0.45\textwidth,height=0.45\textwidth]{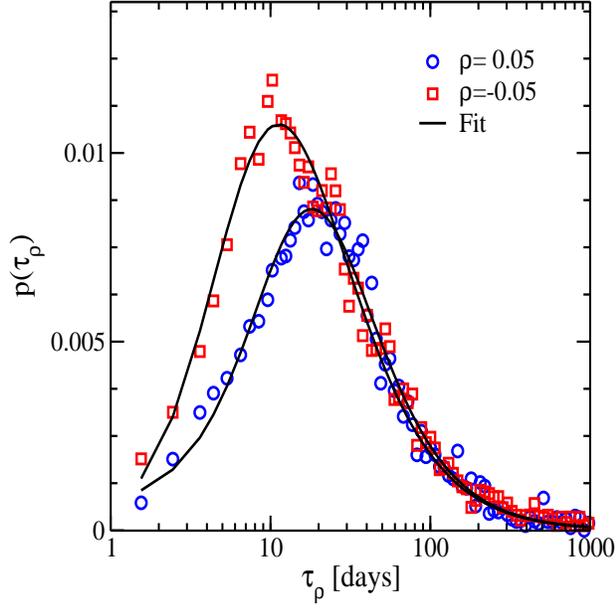}    }\quad
  \subfigure[SP500 (1960.8--2000.3)]{
  \includegraphics*[width=0.45\textwidth,height=0.5\textwidth]{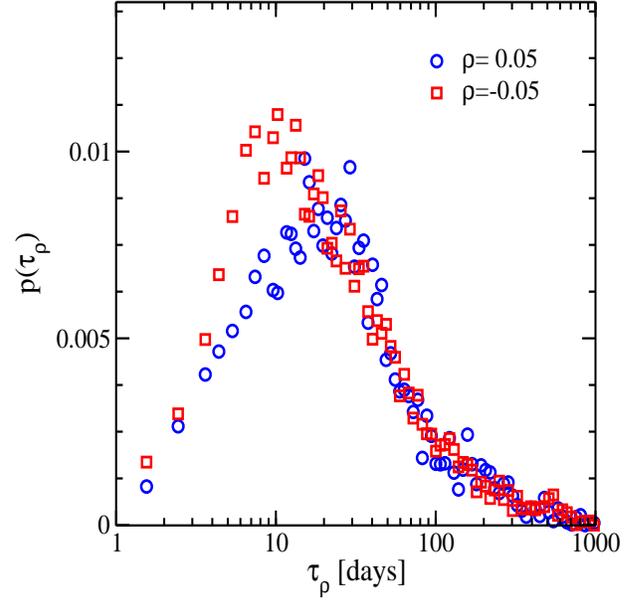} } 

\caption{Inverse statistics for detrended closing prices (open
  symbols) of the SP500 for the time periods indicated. For all cases
  the return levels used were $\left|\rho\right|=0.05$. The solid
  lines represent the best fits of Eq.~(\protect\ref{fit-func}) to
  the empirical data with the following parameters:  (a) SP500
  1940.9--2000.3): $\alpha \approx 0.50$, $\beta \approx 4.87$, $\nu
  \approx 0.88$ and $t_0 \approx 1.59$(loss distribution); $\alpha
  \approx 0.50$, $\beta \approx 5.10$, $\nu \approx 2.56$ and $t_0
  \approx 14.0$ (gain distribution); (b) Same as
  Fig.~\protect\ref{InvDist_SP500}(a), but for a shorter time period
  (1960.8--1999.8).  Note that the tail exponents $\alpha+1$ are very
  close to the ``random walk value'' of $3/2$ for all distributions.
}
   \label{InvDist_SP500} 
\end{figure}

\begin{figure}[t]
  \centering
  \includegraphics*[height=9cm,width=13.5cm]{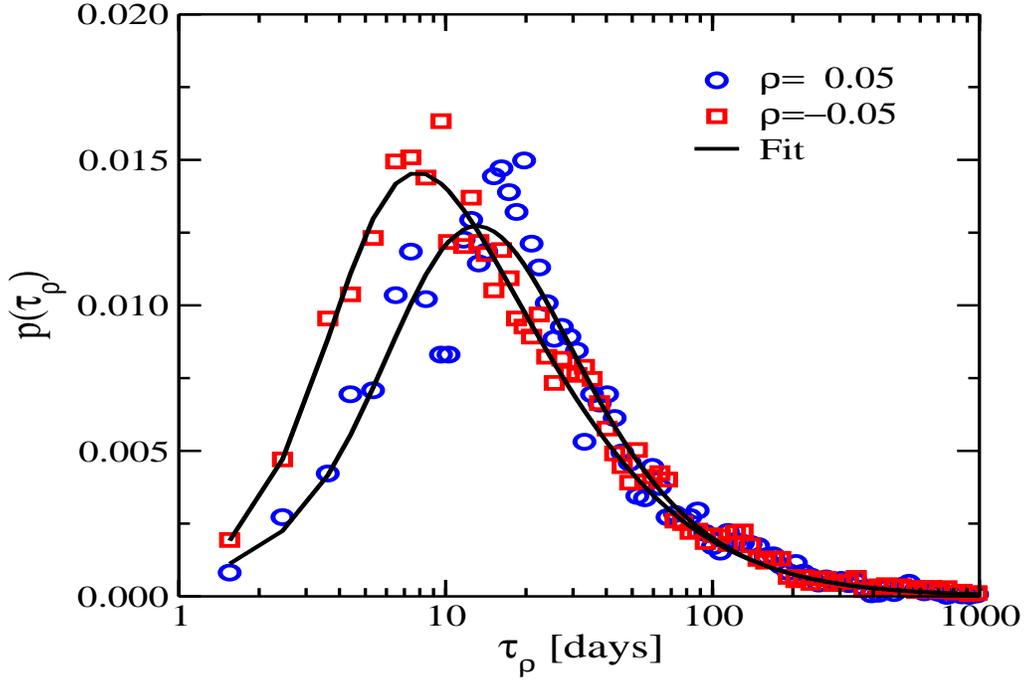}
  \caption{\label{fullnasdaq} Same as
    Fig.~\protect\ref{InvDist_DJIA}(a), 
    but for the Nasdaq. The historical time period considered is
    1971.2 to 2004.6. Again he solid lines represent fit of the
    empirical data against Eq.~(\protect\ref{fit-func}) with
    parameters: $\alpha \approx 0.51$,
    $\beta \approx 4.72$, $\nu \approx 0.73$ and $t_0 \approx 7.92$
    (loss distribution); $\alpha \approx 0.51$, $\beta \approx 4.16$,
    $\nu \approx 2.41$ and $t_0 \approx 0.07$ (gain distribution).
    Note again that the tail exponents $\alpha+1$ are very close to
    the ``random walk value'' of $3/2$ for both distributions.}
\end{figure}

The results so far have been very encouraging with respect to
excellent parametrization of the empirical probability distributions
for three major stock markets, namely DJIA, SP500 and Nasdaq; cf.
Figs.~\ref{InvDist_DJIA}(a), \ref{InvDist_SP500}(a) and
\ref{fullnasdaq} for examples using a return level of $5\%$ and the
figure captions for values of the fit parameters.  The choice of
$\rho=\pm 0.05$ is not accidental. It is sufficiently large to be
above the ``noise level'' quantified by the historical volatility and
sufficiently small to be of quite frequently occurrence. We have also
considered other return levels, showing qualitatively the same
features, but there results are not shown. 

\begin{figure}[t]
  \centering
  \includegraphics*[height=8.5cm,width=13.5cm]{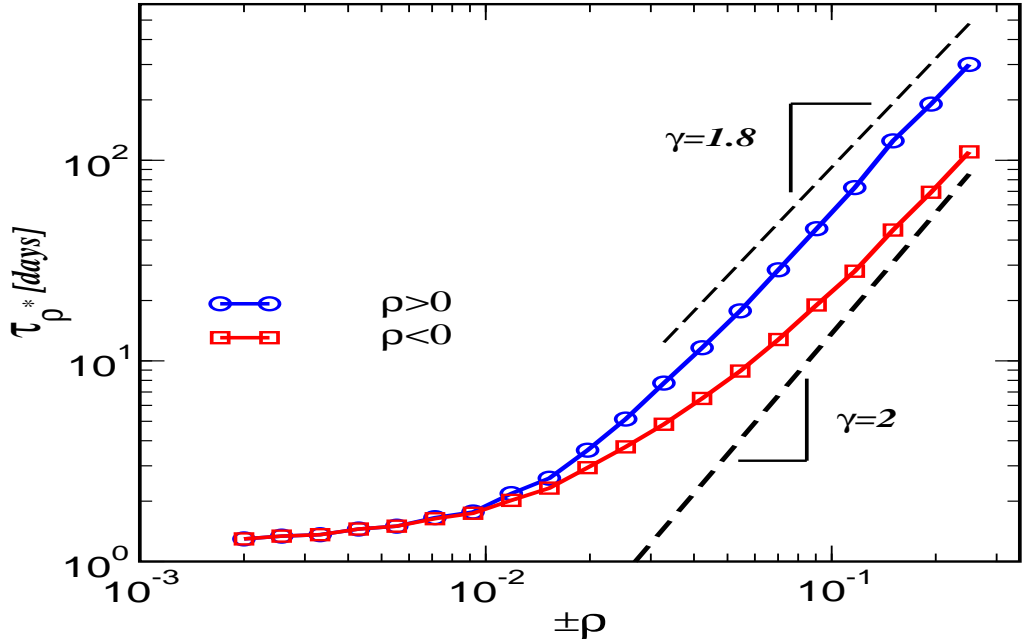}
  \caption{\label{optilevel} The optimal investment horizon
    $\tau^*_\rho$ for positive~(open circles) and negative~(open
    squares) levels of return $\pm \rho$ for the DJIA. In the case
    where $\rho<0$ one has used $-\rho$ on the abscissa for reasons of
    comparison.  If a geometrical Brownian price process is assumed,
    one will have $\tau^*_\rho\sim \rho^\gamma$ with $\gamma=2$ for
    all values of $\rho$. Such a scaling behaviour is indicated by the
    lower dashed line in the graph.  Empirically one finds
    $\gamma\simeq 1.8$~(upper dashed line), only for large values of
    the return.}
\end{figure}

In all three cases, the tail-exponents $\alpha+1$ of the distributions
parameterized by Eq.~(\ref{fit-func}) are indistinguishable from the
``random walk value'' of $3/2$, which is not very surprising.  What
{\it is} both surprising and very interesting is that these three
major US stock markets (DJIA, SP500 and Nasdaq) exhibit a very
distinct gain/loss asymmetry, {\it i.e.}, the distributions are not
invariant to a change of sign in the return $\rho$
(Figs.~\ref{InvDist_DJIA}(a), \ref{InvDist_SP500}(a) and
\ref{fullnasdaq}).  Furthermore, this gain/loss asymmetry quantified
by the optimal investment horizon defined as the peak position of the
distributions has for at least the DJIA a surprisingly simple
asymptotically power law like relationship with the return level
$\rho$, see Fig.~\ref{optilevel}.

See Ref.~\cite{invfx} for an application of inverse statistics to
high-frequency foreign exchange data, specifically the US\$ against
the Deutch Mark and the Japanese Yen. We are also currently
investigating the use of inverse statistics for intra-day stock
prices, but the results so far are still preliminary.

\section{New Results}

As mentioned previously, the purpose of the present paper is to
further investigate the origin of the gain/loss asymmetry in DJIA. We
do that by simply comparing the gain and loss distributions of the
DJIA with the corresponding distributions for a single stocks in the
DJIA as well as their average.

An obvious problem with this approach is that the stocks in the DJIA
changes with historical period and hence an exact correspondence
between the DJIA and the single stocks in the DJIA is difficult to
obtain if one at the same time wants good statistics.  This is the
trade-off, where we have put the emphasis on good statistics. However,
{\it if} this choice produces interpretational difficulties, then one
must ask why analyze the historical DJIA at all?

The 21 company stocks analyzed and presently in the DJIA (by the
change of April 2004) are listed in table \ref{complist} together with
the their date of entry into the DJIA as the time period of the data
set analyzed.

As previously mentioned, the empirical distributions for $\rho=\pm
0.05$ are presented in Figs.~\ref{InvDist_DJIA}(a),
\ref{InvDist_SP500}(a) and \ref{fullnasdaq} for the DJIA, the SP500
and the Nasdaq respectively for the entire time span available to the
authors.  Furthermore, in Figs.~\ref{InvDist_DJIA}(b) and
\ref{InvDist_SP500}(b) we have truncated the full data sets of the
DJIA and SP500 to a shorter historical time period in order to compare
the results for the indices with that of the individual stocks.  What
one should primarily note in the comparison between the two sets of
figures, {\it i.e.}, the longer data sets with the shorter, is that
the only significant difference is to be found in the weaker
statistics of the shorter periods; the positions of the maxima for the
positive and the negative gains are in both cases roughly just below
$10$ and $20$ days, respectively, {\it i.e.}, a difference of roughly
a factor of 2 for all three indices for a return level of $5\%$.

\begin{figure}[t]
  \centering
  \subfigure[Boeing Airways]{
    \includegraphics*[width=0.45\textwidth,height=0.45\textwidth]{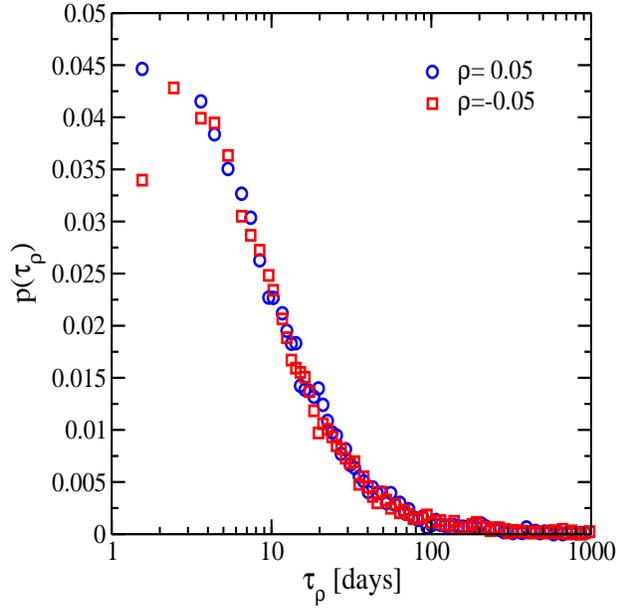} }\quad
  \subfigure[General Electric ]{
    \includegraphics*[width=0.45\textwidth,height=0.45\textwidth]{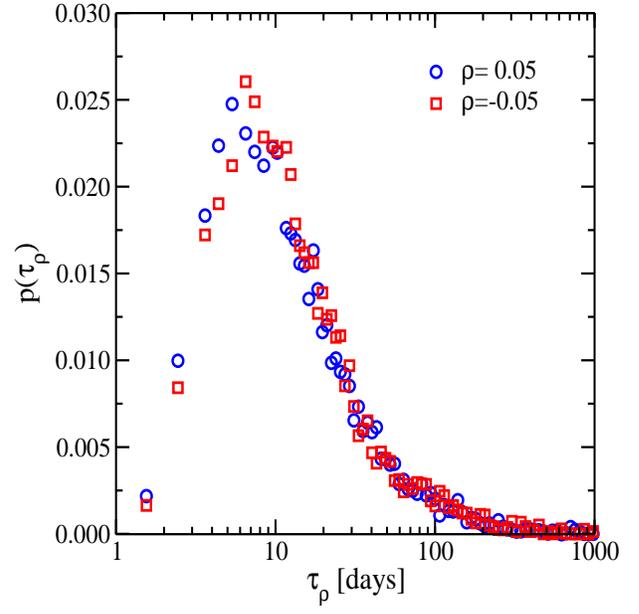} } \\*[0.5cm]
  \subfigure[General Motors ]{
    \includegraphics*[width=0.45\textwidth,height=0.45\textwidth]{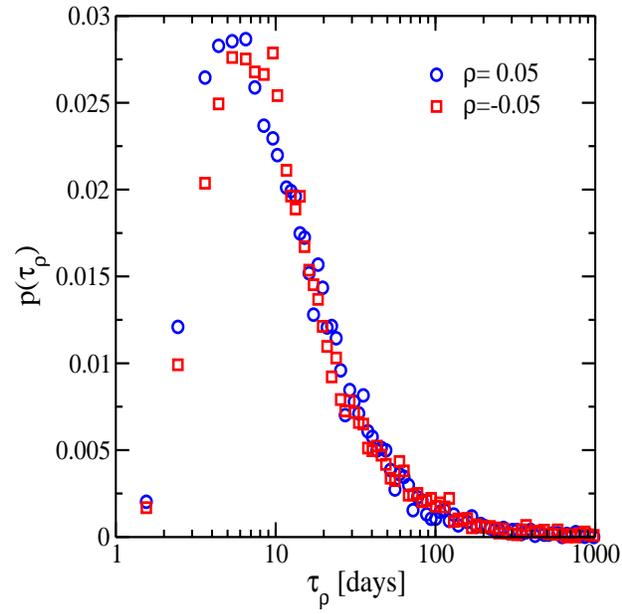}}\quad
  \subfigure[Exxon \& Mobil]{
  \includegraphics*[width=0.45\textwidth,height=0.45\textwidth]{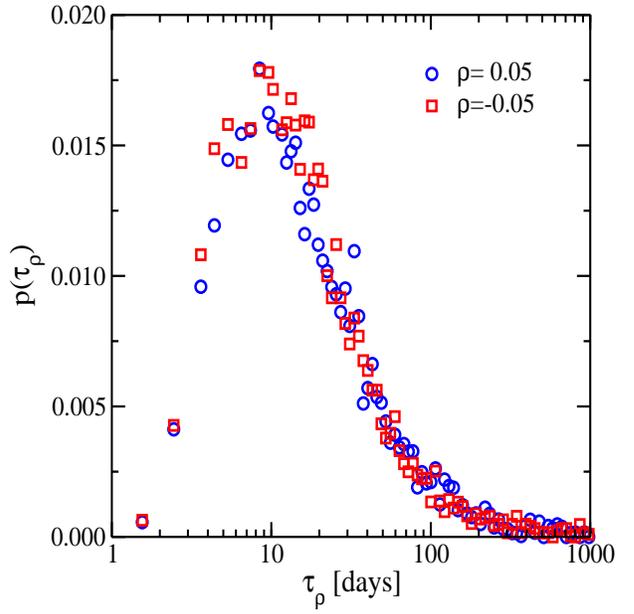} } 

\caption{Same as Fig.~\protect\ref{InvDist_DJIA}(a), but for some of the
  individual companies of the DJIA: (a) Boeing Airways
  (1962.1--1999.8); (b) General
  Electric (1970.0--1999.8); (c) General Motors (1970.0--1999.8);
  (d) Exxon \& Mobil, former Standard Oil (1970.0--1999.8). }
   \label{fig:DJIA-companies}
\end{figure}

In Figs.~\ref{fig:DJIA-companies} 
we show the waiting time distributions for 4 companies in the DJIA,
which are representative for the distributions obtained for all the
companies listed in Table \ref{complist}. We see that, for a return
level $\left|\rho\right|=0.05$, the value of the optimal investment
horizon, {\it i.e.} the position of the peak in the distribution,
ranges from around 2 days to around 10 days depending on the company.
More importantly, it is clear from just looking at the figures that,
within the statistical precision of the data, the distributions are
the same for positive and negative values of $\rho$.

\begin{figure}[t]
  \centering
  \includegraphics*[height=9cm,width=13.5cm]{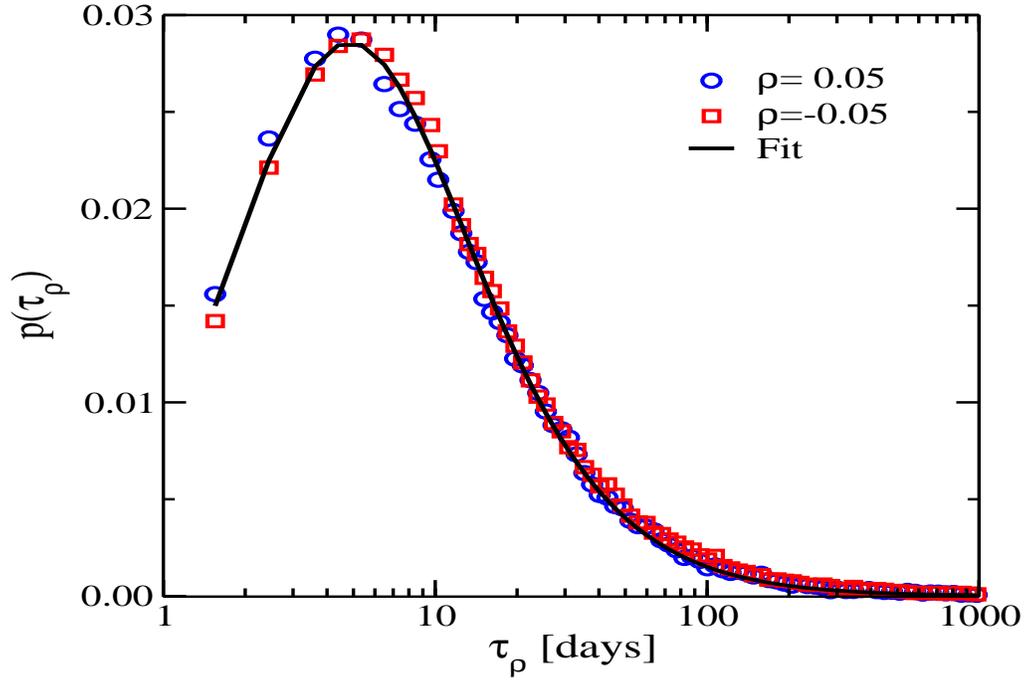}
  \caption{\label{avestock} Averaged gain and loss distribution for
    the companies listed in table \protect\ref{complist}. The fit is
    Eq.~(\protect\ref{fit-func}) with values $\alpha \approx 0.60$,
    $\beta \approx 3.24$, $\nu \approx 0.94$ and $t_0 \approx 1.09$.
    Note that the tail exponent $\alpha+1$ is $0.1$ above the ``random
    walk value'' of $3/2$.}
\end{figure}

In order to further quantify this invariance with respect to the sign
of $\rho$, we have averaged the (company) gain and loss distributions
separately in order to obtain an average behavior for the stocks
listed in Table~\ref{complist}.  The result of this averaging process
(Fig.~\ref{avestock}) is nothing less that an almost perfect agreement
between the gain and loss distributions with a peak position around 5
days for both distributions.  This means that the optimal investment
horizon for the average of these selected individual stocks is
approximately half that of the loss distribution for the DJIA and
approximately one fourth of that for the gain distribution. In other
words, it is twice as slow to move the DJIA down and four times as
slow to move the DJIA up compared to the average time to move an
individual stock in the DJIA up or down.

How can we rationalize these results? What we have done in essence is
to interchange the operations of averaging over the stocks in the DJIA
and calculating the inverse statistics for the stocks of this index.
Since the DJIA is constructed such that it covers all sectors of the
economy of which it seems quite reasonable to assume that a $5\%$
gain/loss in the shares of for example Boeing Airways in general has
nothing fundamentally to do with a corresponding gain/loss in the
shares of Coca-Cola Company {\it especially} since the data are
detrended. In other words, it seems quite reasonable to assume that
there is nothing special about a return level of $5\%$ in terms of
economic fundamentals {\it etc.} This assumption is also strongly
supported by the results presented in Fig.~\ref{optilevel}.

This then means that the two operations, {\it i.e.} the averaging and
the inverse statistics calculation, do not commute not even
approximately. Hence significant inter-stock correlations
must exist even for a rather modest return level $\rho = 0.05$.
In our view, this is a quite surprising result and especially
considering that the large differences in the optimal investment
horizons for the distributions of the index and the average of the
individual stocks.

\section{Summary and Conclusions}

We have considered inverse statistics for financial data. It is argued
that the natural candidate for such statistics is what we call the
investment horizon distribution. Such a distribution, obtained from
the historic data of a given market, indicates the time span an
investor historically has to wait in order to obtain a predefined
level of return. For the three major US markets, namely the DJIA,
SP500 and Nasdaq, the distributions are parametrized excellently by a
shifted generalized Gamma distributions for which the first moment
does not exist.

The typical waiting time, for a given level of return $\rho$, can
therefore be characterized by {\it e.g.} the time position of the
maximum of the distribution which we call the {\em optimal} investment
horizon. By studying the behaviour of this quantity for
positive~(gain) and negative~(loss) levels of return, a very
interesting and pronounced gain/loss asymmetry emerges for all three
markets considered.  The nature of this gain/loss asymmetry was
further investigated by calculating the investment horizon
distribution for $21$ of the  $30$ individual stocks in the DJIA. Quite
surprisingly the observed gain/loss asymmetry in the DJIA is {\it not
  present} in the investment horizon distribution for the individual
stocks.

Specifically, we have shown that if one first ``average'' the stock
prices of, say, the DJIA in order to create the index and then
calculate the inverse statistics of it, then one obtains a pronounced
gain/loss asymmetry.  Reversing the order of these two operations,
however, makes this asymmetry {\it disappear}. Admittedly, this has
been done systematically for only a single gain and loss level $\rho$,
but it seems unreasonable to claim anything special about the 5\%
gain/loss level used. Hence, the investment horizon distribution for
the individual stocks is {\em invariant} under a change of sign for
$\rho$.

Furthermore, the optimal investment horizon for the average of the
distributions for the individual stocks is approximately half that of
the loss distribution for the entire market (DJIA) and approximately
one fourth of that for the gain distribution for the entire market. In
other words, it is twice as slow to move the DJIA down and four times
as slow to move the DJIA up compared to the average time to move an
individual stock in the DJIA up or down.

There are several possible scenarios which may explain the observed
behavior. However, they all amount to more or less the same thing. A
down/up-turn in the DJIA may be initiated by a down/up-turn in some
particular stock in some particular economical sector. This is
followed by a down/up-turn in economically related stocks in other
sectors and so forth. The result is a cascade, or synchronization,
of consecutive down/up-turns in all the sectors covered by the DJIA.
The initiation of this may be some more general new piece of
information, which is considered more crucial for one sector than
other sectors, but as argued for in length in~\cite{Schiller} it may
also happen for no obvious reason what so ever.

An (rational) example would be that Intel drops significantly due to
bad quarterly earnings in turn, by a cascade process, affecting the
stock price of IBM and MicroSoft and so forth. As the index, at least
from a physicist's point of view, can be compared to an external
field, movements in the index due to a single or a few stocks can
rapidly spread through most or all sectors, {\it if} psychology in
general and specifically feed-back loops are important, thus creating
a relatively large movements in the over-all index. That losses in
general are faster than gains must also be attributed to human
psychology: people are in general much more risk adverse than risk
taking\footnote{The reason for this has an evolutionary origin: one of
  the functions of emotions is to speed up decisions, which is an
  excellent feature in terms of physical survival. However, in the
  case of the stock markets, where financial survival may be at stake,
  rational decisions are to be prefered. But, as rational decisions
  take longer time to make, people will often act on emotions. In a
  study, published in June in the journal Psychological Science by a
  team of researchers from Stanford University, Carnegie Mellon
  University, and the University of Iowa, a group of functional
  psychopats outperformed a group of people with normal emotional
  responses in a financial game.}.

In conclusion, the results presented here thus provide further
evidence for the presence of cooperative behavior (or synchronization)
possibly with a psychological origin in the stock market beyond what
can be deduced from standard financial models.

\section*{Acknowledgment}

I.S.\ would like to thank the Nordic Institute for Theoretical Physics
(NORDITA) for kind hospitality where the last part of this work was
completed.




\newpage

%

\begin{table}
\centering
\begin{tabular}{|l||ll|} 
 \hline
 Company  & Entering date  &  Data period \\
 \hline \hline
Alcoa$^\star$           &  Apr 22, 1959   & 1962.1--1999.8  \\
American Express Company& Aug 30, 1982   & 1977.2--1999.8 \\
ATT$^\dagger$           &  Mar 14, 1939   & 1984.1--1999.8 \\
Boeing Airways          &  Jul 08, 1986   & 1962.1--1999.8 \\
Citicorp$^\bullet$      &  Mar 17, 1997   & 1977.0--1999.8 \\
Coca-Cola Company       &  Mar 12, 1987   & 1970.0--1999.8\\
DuPont                  &  Nov 20, 1935   & 1962.1--1999.8\\
Exxon \& Mobil$^\circ$  & Oct 01, 1928    & 1970.0--1999.8  \\
General Electric        &  Nov 07, 1907   & 1970.0--1999.8\\
General Motors          &  Mar 16, 1915   & 1970.0--1999.8\\
Goodyear                &  July 18 1930   & 1970.1--1999.8\\
Hewlett \& Packard      &  Mar 17, 1997   & 1977.0--1999.8\\
IBM                     &  Jun 29, 1979   & 1962.0--1999.8\\ 
Intel                   &  Nov 01, 1999   & 1986.5--1999.8\\
International Paper     &  Jul 03, 1956   & 1970.1--1999.8\\
Eastman Kodak Company   &  Jul 18, 1930   & 1962.0--1999.8\\
McDonald's Cooperation  &  Oct 30, 1985   & 1970.1--1999.8\\
Merck \& Company        &  Jun 29, 1979   & 1970.0--1999.8\\
Procter \& Gamble       &  May 26, 1932   & 1970.0--1999.8\\
The Walt Disney Co.     &  May 06, 1991   & 1962.0--1999.8\\
Wall Mart               &  Mar 17, 1997   & 1972.7--1999.8\\
\hline 
\end{tabular}
\caption{\label{complist} List of the ($21$) DJIA stocks analyzed in
  this work (about $70\%$ of the total number). 
  Furthermore, their date of entry into the DJIA are shown, and  the
  time period covered by the analyzed data set. All of 
  these companies are also naturally part of SP500 with General Electric as the 
  most heavily weighted stock.$^\star$Former Aluminum Corporation of America.
  $^\dagger$Former American Tel. \& Tel. Truncated due to anti-trust case in 
  1984.  $^\bullet$Former Travelers Group. $^\circ$Former Standard Oil.}
\end{table}

%

\newpage

\newpage

\newpage

\newpage

\newpage

\newpage

\newpage 

\end{document}